\documentstyle[prl,aps,epsf,epsfig,floats,twocolumn]{revtex}

\begin{document}
\draft

\twocolumn[\hsize\textwidth\columnwidth\hsize\csname
@twocolumnfalse\endcsname

\title{
\hfill{\small{DOE/ER/40762-231}}\\
\hfill{\small{UMD-PP\# 01-056}}\\[0.6cm]
Is the Sullivan Process Compatible with QCD Chiral Dynamics?}
\author{Jiunn-Wei Chen and Xiangdong Ji}
\address{Department of Physics, University of Maryland,
College Park, MD 20742-4111}
\maketitle
\begin{abstract}
We calculate the leading non-analytic quark-mass 
dependence in the moments of isovector 
quark distributions using heavy-baryon chiral 
perturbation theory. The results differ from what has 
been obtained from the Sullivan process in which hard scattering
occurs through the virtual pion cloud of the nucleon.
Our results provide useful guidance in formulating
meson-cloud models consistent with chiral dynamics 
and can be used to constrain the extrapolations of the existing 
lattice QCD results to the physical quark masses
\end{abstract}
\medskip ]

The nucleon (proton and neutron) has a quark sea. For 
a long time, the leading picture for this sea comes from 
the meson-cloud model in which the nucleon can 
virtually dissociate into mesons such as pions, 
kaons, etc., plus a baryon core. 
A concrete realization of this model can be found in 
lepton-nucleon deep-inelastic scattering in which 
the (Bjorken-like) virtual photon scatters deep-inelastically 
off the meson cloud. The Feynman diagram for this process 
is shown in Fig. 1 and is called the Sullivan process 
\cite{sullivan}. Many
interesting results from the meson-cloud model have been
obtained, which have shed important light 
on the dynamics of the quark sea \cite{thomas,others,review}. 

\begin{figure}[h]
\begin{center}
\epsfxsize=4.5cm
\centerline{\epsffile{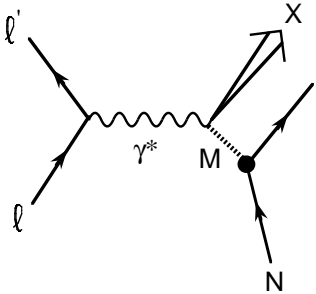}}
\end{center}
\caption{The Sullivan process in which the virtual photon
scatters off the meson cloud in the nucleon.}
\end{figure}

A crucial question about the meson-cloud model and
the associated Sullivan process is to what extent the model
can be justified from the underlying quantum chromodynamics 
(QCD) and its low-energy
effective theories? Recently, A. Thomas et. al.
pointed out the best place to check is the chiral properties
of the model predictions \cite{thomas1}. They worked out, 
for the first time, the detailed leading non-analytic quark
(pion) mass dependence of the parton distributions probed
in the Sullivan process. This letter seeks to answer 
the above question by calculating the same type of
chiral logarithms using the low-energy effective theory of QCD: 
the heavy-baryon chiral perturbation theory \cite{HBChPT}. 
As it turns out, there are subtle differences between 
the traditional approach in calculating
the Sullivan process and chiral perturbation theory. 
Our results can be used to improve upon the 
formulation of the Sullivan process and to study the 
quark mass dependence of the lattice QCD calculations. 

We are interested in the moments of the quark distributions 
in the proton, 
\begin{eqnarray}
\left\langle x^{n-1}\right\rangle _{q} &=&\int_{0}^{1}dxx^{n-1}\left(
q\left( x\right) +\left( -1\right) ^{n}\overline{q}\left(
x \right) \right)  \ , 
     \nonumber \\
\left\langle x^{n-1}\right\rangle _{\Delta q} &=&\int_{0}^{1}dxx^{n-1}\left(
\Delta q\left( x\right) +\left( -1\right) ^{n-1}\Delta \overline{q%
}\left( x\right) \right) \ , 
    \nonumber \\
\left\langle x^{n-1}\right\rangle _{\delta q} &=&\int_{0}^{1}dxx^{n-1}\left(
\delta q\left( x\right) +\left( -1\right) ^{n}\delta \overline{q%
}\left( x\right) \right) \ , 
    \nonumber \\
 &&~~~~~~~~~(n=1,2,3,...) 
\end{eqnarray}
where $q$ ($\overline{q}$) is the quark (antiquark) spin-averaged
distribution, $\Delta q$ $(\Delta \overline{q})$ the 
helicity distribution, and $\delta q$ $(\delta \overline{q})$
the transversity distribution \cite{filippone}. 
The variable (Feynman) $x$ 
is the momentum fraction of the proton carried by a quark
in the infinite momentum frame, and for simplicity 
we have suppressed 
the renormalization scale dependence. In QCD, 
the moments are related to the forward matrix elements 
of the twist-two operators
\begin{eqnarray}
{\cal O}_{q}^{\mu _{1}\cdots \mu _{n}} &=&\overline{q}\gamma ^{(\mu
_{1}}iD^{\mu _{2}}\cdots iD^{\mu _{n})}q \ ,  \nonumber \\
\widetilde{{\cal O}}_{q}^{\mu _{1}\cdots \mu _{n}} &=&\overline{q}\gamma
^{(\mu _{1}}\gamma _{5}iD^{\mu _{2}}\cdots iD^{\mu _{n})}q  \ , 
\nonumber \\
\widetilde{{\cal O}}_{Tq}^{\alpha\mu_1\cdots \mu _{n}} 
&=&\overline{q}\sigma
^{\alpha(\mu_1}\gamma_5 iD^{\mu _{2}}\cdots iD^{\mu _{n})}q  \ , 
\label{op}
\end{eqnarray}
through the relations, 
\begin{eqnarray}
\left\langle PS\left| {\cal O}_{q}^{\mu _{1}\cdots \mu
_{n}}\right| PS\right\rangle &=&
 2\left\langle x^{n-1}\right\rangle_{q}
   P^{(\mu _{1}}\cdots P^{\mu _{n})} \ ,  \nonumber \\
\left\langle PS\left| \widetilde{{\cal O}}%
_{q}^{\mu _{1}\cdots \mu _{n}}\right| PS\right\rangle
  &=& 2\left\langle x^{n-1}\right\rangle _{\Delta q}
  MS^{(\mu _{1}}P^{\mu _{2}}\cdots P^{\mu
_{n})}   \ , 
 \nonumber \\
\left\langle PS\left| \widetilde{\cal O}
_{Tq}^{\alpha\mu_1\cdots \mu _{n}}\right| PS\right\rangle
  &=&  \nonumber 
2\left\langle x^{n-1}\right\rangle_{\delta q} \nonumber \\
&& \times
  MS^{[\alpha}P^{(\mu_1]}P^{\mu _{2}}\cdots P^{\mu
_{n})}  \ ,
\end{eqnarray}
where $(\cdots)$ and $[\cdots]$ denote, respectively, 
the symmetrization and antisymmetrization of the indices
in between. $|PS\rangle$ is the ground state of the 
nucleon with four-momentum $P^\mu$ and 
polarization vector $S^\mu$ ($S^2=-1$). $M$ is the nucleon
mass. All tensors are trace free.

To calculate the leading non-analytic quark-mass dependence 
of these moments in heavy-baryon chiral perturbation theory \cite{HBChPT}, we need to construct a chiral expansion 
of the quark operators in terms of the hadronic (pion and
nucleon) operators with the identical symmetry properties. 
In this study, we focus on the isovector combinations
of the quark densities, up minus down quarks.
To leading order in chiral power counting, we
have,
\begin{eqnarray}
{\cal O}_{u-d}^{\mu _{1}\cdots \mu _{n}} 
  &\sim&  \overline{N} v^{(\mu
_{1}}\cdots v^{\mu _{n})} \left(u\tau _{3}u^\dagger + 
  u^\dagger \tau_3 u
   \right)N  \nonumber \\
&&        + \alpha \overline{N}S^{(\mu _{1}}v^{\mu _{2}}\cdots 
v^{\mu_{n})} \left(u^\dagger\tau _{3} u - u 
   \tau_3 u^\dagger\right)N
\ ,  \nonumber \\
\widetilde{{\cal O}}_{\Delta u-\Delta d}^{\mu _{1}\cdots \mu _{n}} 
&\sim&
  \overline{N}S^{(\mu
_{1}}v^{\mu_2}\cdots v^{\mu _{n})} 
\left(u\tau _{3}u^\dagger + u^\dagger \tau_3 u
   \right) N          \nonumber \\
&&        + \beta \overline{N}v^{(\mu _{1}}\cdots v^{\mu
_{n})} \left(u^\dagger\tau _{3} u - u \tau_3 u^\dagger\right)N
\ ,   \\
\widetilde{{\cal O}}_{\delta u-\delta d}^{\mu _{1}\cdots \mu _{n}} 
&\sim& \overline{N}S^{[\alpha}v^{(\mu_1]}\cdots v^{\mu _{n})} \left(u^\dagger\tau _{3}u^\dagger + u \tau_3 u
   \right)N   \nonumber \\
&&  + \gamma \overline{N}S^{[\alpha}S^{(\mu_1]}v^{\mu_2}\cdots v^{\mu
_{n})} \left(u^\dagger\tau _{3} u^\dagger - u \tau_3 u \right)N
\nonumber
\label{op2}
\end{eqnarray}
where $N$ and ${\overline N}$ are the nucleon fields, 
$v^{\mu }$ is the nucleon four-velocity, and 
$u = \exp(i\pi^a\tau^a/2f_\pi)$ with pion fields $\pi^a$
and decay constant $f_{\pi }=93$ MeV. $\alpha$, $\beta$
and $\gamma$ are unknown coefficients. For $n=1$,
there is also a leading pion operator in 
${\cal O}^{\mu_1}_{u-d}$, which has not been shown 
explicitly.

\begin{figure}[h]
\begin{center}
\epsfxsize=7.5cm
\centerline{\epsffile{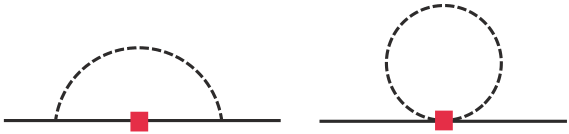}}
\end{center}
\caption{Feynman diagrams contributing to the leading
chiral logarithms in the nucleon matrix elements of twist-two
operators. The dashed lines represent pions.}
\end{figure}

The contributions to the leading chiral logarithms 
of the twist-two matrix elements come 
from three separate sources: first, the wave function 
renormalization of the nucleon states; second, 
the virtual-pion cloud as
shown in the first diagram in Fig. 2; third, the 
pion tad-pole contribution as shown in the second
digram in Fig. 2. 
A straightforward calculation yields,
\begin{eqnarray}
\left\langle x^{n}\right\rangle _{u-d} &=&C_{n}\left\{ 1-\frac{\left(
3g_{A}^{2}+1\right) m_{\pi }^{2}}{(4\pi f_{\pi })^2}\ln \left( \frac{%
m_{\pi }^{2}}{\mu ^{2}}\right) \right\} \ ,  \label{mom1}\nonumber \\
\left\langle x^{n-1}\right\rangle _{\Delta u-\Delta d} &=&\widetilde{C}_{n-1}%
\left\{ 1-\frac{\left( 2g_{A}^{2}+1\right) m_{\pi }^{2}}
{(4\pi f_{\pi
})^{2}}\ln \left( \frac{m_{\pi }^{2}}{\mu ^{2}}\right) \right\} \ ,\nonumber \\
\left\langle x^{n-1}\right\rangle _{\delta u-\delta d} 
&=&\overline{C}_{n-1}%
\left\{ 1-\frac{\left( 4g_{A}^{2}+1\right) m_{\pi }^{2}}
   {2(4\pi f_{\pi})^{2}}\ln \left( \frac{m_{\pi }^{2}}{\mu ^{2}}\right) \right\} \ , \nonumber 
 \\
n &=&1,2,\cdots  
\end{eqnarray}
where $C_n$, $\tilde C_n$, and $\overline{C}_n$
are the corresponding moments in the chiral limit, $g_{A}$ is the 
isovector axial charge in the same limit, 
and the renormalization scale $\mu $ can be taken to be
$4\pi f_{\pi }\sim 1$ GeV. 
The relative coefficients between the results
in the chiral limit and the leading chiral-logarithms 
are seen to be fixed. 
For $\left\langle 1\right\rangle _{\Delta u-\Delta
d}$, which is the physical isovector axial charge, 
our leading non-analytic correction 
agrees with the previous calculation \cite{BKM}. 
$\left\langle 1\right\rangle _{u-d}$ is the
isospin charge which is protected from chiral corrections
by the isospin symmetry.  
The next important chiral correction 
is of type $m_{\pi }^{2}/(4\pi f_\pi)^2$ 
which is linear and analytical in quark masses.
These corrections cannot be computed without 
additional parameters. 

The leading non-analytical quark mass dependence of
$\left\langle x^{n}\right\rangle _{u-d}$ 
was calculated using the pion-cloud
model (the Sullivan process) in ref.\cite{thomas1}. 
Instead of the pre-factor $\left(3g_{A}^{2}+1\right)$ 
in front of the chiral logarithm, they found $4g_{A}^{2}$. 
Although in the $g_A=1$ limit the two results agree,
they are different in general. In particular, in the
limit of a large number of colors ($N_c$), 
$g_A\sim N_c$ \cite{manohar}. 

We find that the above discrepancy comes from 
the differences between the linear and non-linear 
formulations of chiral expansion. While the 
non-linear formulation used in our calculation
has a simple and clear power counting scheme, the linear
formulation does not. The Sullivan process is 
traditionally done in the linear sigma model with the
following lagrangian 
\begin{equation}
  {\cal L}_{\rm linear-\sigma}
   = \bar \Psi \left(i\not\!\partial
  + g_{\pi NN}(\sigma + 
i\vec{\tau}\cdot\vec{\pi}\gamma_5)\right) \Psi
         + ...
\label{linearsig}
\end{equation}
where $\Psi$ represents the nucleon field in the linear 
representation and $g_{\pi NN}$ is the nucleon-pion coupling. 
The pure meson sector of the lagrangian has been omitted. 
After spontaneous symmetry
breaking, the above model in principle
yields a Goldberg-Treimann relation 
$g_{\pi NN} = M/f_\pi$ with $g_A=1$. In the literature 
\cite{thomas1}, however, $g_{\pi NN}$ in the 
linear-$\sigma$ model result is usually replaced by the 
full Goldberg-Treimann relation
$g_{\pi NN} = g_A M/f_\pi$. 

To make the linear $\sigma$ model result consistent
with that of non-linear chiral expansion, 
one must add higher-dimensional
operators to the lagrangian in Eq.~(\ref{linearsig}). 
As we have mentioned above, the interaction terms of
different mass dimensions contribute to the same order 
of chiral power is the main disadvantage 
of the linear formulation. In the present example,
we need to add the following dimension-five term
\cite{weinberg},
\begin{eqnarray}
  {\cal L}' &=& g'\overline{\Psi}
\left( (\vec{\tau}\cdot\vec{\pi}i\not\!\partial \sigma
  - \sigma\vec{\tau}\cdot i\not\!\partial\cdot \vec{\pi})\gamma_5
       \right.   \nonumber \\
 && \left. + \vec{\tau}\cdot(\vec{\pi}
\times i\not\!\partial\vec{\pi})
\right)\Psi \ . 
\end{eqnarray}
To restore the full Goldberger-Treiman relation, 
$g'\sim g_A-1$. Setting $g_A=1$ in the result
of Ref. \cite{thomas1} and adding the 
additional contributions generated from the above 
lagrangian, we find the chiral
logarithms in the linear formulation
coincide completely with the results above.

Our results can be used to constrain the
quark mass dependence of the matrix elements
calculated in lattice QCD \cite{thomas1,dolgov}. 
Indeed the isovector combinations of the moments are
easier to obtain since they do not require a 
computation of expansive ``disconnected'' diagrams. 
The results can also be used to improve the 
meson-cloud model to the point where it becomes
consistent with the chiral physics of QCD.

To summarize, we have calculated the leading non-analytic
quark mass dependence of the twist-two, isovector quark 
matrix elements in the nucleon. The results indicate 
that the traditional approach in calculating the 
Sullivan process needs to be modified to be 
fully consistent with the chiral dynamics of QCD.

Note added in proof: After this work was completed, 
we learned a similar work by D. Arndt and M. J. Savage
\cite{AS}. For the same quantities computed, 
both papers agree.

\acknowledgements
We thank W. Melnithouk and A. Thomas for useful
conversations and correspondences. This work is 
supported in part by the U.S. Dept. of Energy 
under grant No. DE-FG02-93ER-40762.

\end{document}